\documentclass[twocolumn,preprintnumbers,amsmath,amssymb]{revtex4-1}
\usepackage{epsfig}
\usepackage{array}
\usepackage{multirow}
\usepackage{color}
\usepackage{float}
\usepackage[font=footnotesize,format=plain,labelsep=period,justification=raggedright]{caption}
\providecommand{\e}[1]{\ensuremath{\times 10^{#1}}}

\begin{document}

\title{Computer simulation study of surface wave dynamics at the crystal--melt
interface.}
\author{Jorge Benet, Luis G. MacDowell and Eduardo Sanz}
\affiliation{Departamento de Qu\'{\i}mica F\'{\i}sica,
Facultad de Ciencias Qu\'{\i}micas, Universidad Complutense de Madrid,
28040 Madrid, Spain}
\date{\today}

\begin{abstract}

We study, by means of computer simulations, the crystal-melt interface of three
different systems: hard-spheres,  Lennard Jones and the TIP4P/2005 water model.
In particular, we focus on the dynamics of surface waves. 
We observe that the processes involved in the relaxation of surface waves are
characterized by distinct time scales: 
a slow one related to the continuous recrystallization and melting, 
that is governed by capillary forces; 
and a fast one 
which we suggest to be due to a combination of processes
that quickly cause small perturbations
to the shape of the interface (like e. g. Rayleigh waves, subdiffusion, or 
attachment/detachment of particles to/from the crystal). 
The relaxation of surface waves becomes dominated by the slow process 
as the wavelength increases.  
Moreover, we see that the slow relaxation is not influenced by \textcolor{black}{the details of the microscopic dynamics}. 
In a time scale characteristic for the diffusion of the liquid phase,  
the relaxation dynamics of the
crystal-melt interface of water is around one order of magnitude slower than that
of Lennard Jones or hard spheres, which we ascribe to the presence of orientational degrees of
freedom in the water molecule.  Finally, we estimate the rate of crystal growth
from our analysis of the capillary wave dynamics and compare it with previous
simulation studies and with experiments for the case of water.

\end{abstract}

\maketitle
\renewcommand{\arraystretch}{1.5}

\section{Introduction}

The crystal-melt interface (CMI) has a great relevance in 
materials science given that its properties have a strong influence in crystal nucleation
and growth, as well as on wetting phenomena \cite{Boettinger00,Woodruff}.  
Despite its importance, the CMI is far less understood than the
fluid-fluid interface because the former is difficult to probe 
with standard
experimental techniques like X-ray diffraction \cite{Adamson}. 
For example, it is well known that the interfacial tension of liquid water
at ambient conditions is 72 mN/m, whereas the reported values for the
ice-water interfacial free energy at ambient pressure range from to 25 to 35 mN/m \cite{pruppacher1995}.
Also the dynamics of the fluid-fluid interface is far better understood 
than that of the CMI \cite{tejero85,jeng98}. 
Understanding the dynamics of the CMI is of great interest given that it can provide 
valuable insight to 
the important process of crystal-growth \cite{karma93,hoyt02}. 

The CMI interface is not flat, but rather exhibits relatively large
undulations of the local interface position, or surface waves (SW), 
as a result of thermal excitations (see online movie \cite{movie}).  
For length scales below the capillary length, 
SW are mainly governed by
the interfacial stiffness and are known under the name of capillary waves (CW).
The equilibrium and dynamic properties of CW at the fluid--fluid
interface have a long history and
were already studied by Smoluchowski and Kelvin \cite{Smoluchowski,landau91}.
For the CMI, the study of the CW spectrum provides
static properties, like the interfacial stiffness or 
the interfacial free energy \cite{PhysRevLett.86.5530,Weeks09}. 

At smaller length scales and higher frequencies, the 
surface of elastic media exhibit thermal excitations
known under the name of Rayleigh waves \cite{landau69}. These are
small amplitude, high frequency perturbations that result from the elastic 
response of the solid. 

Rayleigh and CW serve as a benchmark for the study of
other surface phenomena in more complex materials. For example, polymer
solutions and
polymer gels, which are able to support both elastic and viscous response,
exhibit a crossover from capillary to elastic SW
\cite{cao91,dorshow93,monroy98}. The CMI  
also appears as an interesting system for the observation of 
SW.  The solid phase is
elastic and could in principle exhibit Rayleigh waves, while the
fluid phase is viscous and could rather favour CW.

Unfortunately, despite the fair amount of theoretical research in the
field, there seems to be no appropriate theoretical framework
for the study of crystal-melt SW dynamics. Pleiner, Harden and
Pincus extended the Rayleigh
theory in order to incorporate the viscoelastic response of polymeric
materials, but did not consider polymers in contact with a viscous dense
phase \cite{pleiner88,harden91}. The theory was later extended
to study a dense fluid on an elastic medium, but capillary forces were
neglected \cite{kumaran95}.  On the contrary, Jeng et al extended the Kelvin
theory to study
SW at the interface of two dense fluids, but did not
incorporate the elastic response of the solid \cite{jeng98}. A suitable
theoretical framework is in principle that of Baus and Tejero, who
considered SW at the interface of two fluids baring
simultaneously viscous and elastic response. However, the final
results were worked out only for the special case of a vapour-liquid interface
where one of the phases has negligible viscosity \cite{baus83,tejero85}.

An alternative rather different approach to study interfacial fluctuations of
the crystal-melt interface is employed in the field of crystal growth. The
emphasis here is on the hydrodynamic equations of heat and mass transport,
and energy dissipation is enforced by introducing gaussian random  noise. In the
limit of small temperature gradients, this formalism provides a diffusion
equation for the interface height fluctuations, and hence a strongly damped interface
dynamics \cite{karma93,hoyt02,amini06}.

Computer simulations (see e.g. Ref. \cite{Hoyt20101382} and references therein)  and experiments of colloidal
systems \cite{Weeks09}
have also been used to investigate the CMI.  
Both approaches allow for the
visualization of the CMI at a single-particle scale. 
Therefore, these techniques are highly suited to improve
our understanding of the CMI. 
Many studies are devoted to obtain relevant static properties
of the interface, such as the stiffness or the interfacial free energy, by means 
of an analysis of the spectrum of interface fluctuations
(e.g. \cite{PhysRevLett.86.5530,Hoyt20101382,Davidchack06,0295-5075-93-2-26006,jcp_aleksandar_yukawa,Oettel12}). 
 \textcolor{black}{The interfacial free energy can also be obtained by other methods like thermodynamic integration 
\cite{broughton:5759,PhysRevLett.100.036104,doi:10.1021/ct300193e,davidchack010,davidchack:7651}, metadynamics \cite{PhysRevB.81.125416}, 
or by combining classical nucleation theory with simulations \cite{bai:124707,knott12,jacs2013}}.
The study of dynamic properties of the CMI, by contrast, has not received that much attention. 
The dynamics of the CMI has only been investigated by means of computer simulations for
metal models \cite{hoyt02} and hard spheres \cite{amini06}, and experimentally
for colloidal suspensions \cite{Weeks09}. Not only there are just a handful of
works dealing with the dynamics of the CMI but the results are in some cases contradictory.
For instance, in simulation studies \cite{amini06} a quadratic dependency of 
the relaxation frequency with the wave vector is observed whereas 
the experimental work of Ref. \cite{Weeks09} claims that such dependency is instead 
linear.

Motivated by the importance of gaining a deeper understanding on
the dynamics of the CMI we pursue in this paper a 
computer simulation study of the relaxation of SW for the CMI of three
archetypal systems, namely hard spheres (HS), Lennard Jones (LJ) and
water. 
We show that the relaxation of crystal-melt SW is 
well described by a double exponential given that
there are different processes, characterised by different
time scales, involved in such relaxation. As the wavelength increases
only one process, the relaxation of CW, is observed. 
We also show that \textcolor{black}{the details of the microscopic dynamics are not important for the relaxation of crystal-melt CW}. 
Moreover, we compare 
the relaxation dynamics of SW for systems composed of molecules
with (water) and without (LJ and HS) orientational degrees of freedom. 
Finally, following the methodology proposed in Refs. \cite{karma93,hoyt02}, we estimate
the kinetic coefficient (the proportionality constant between the speed of
crystal growth and the supercooling) from our measurements of the CW relaxation
dynamics, and compare our results with independent measurements of such parameter.

\section{Methods}

We simulate a solid in equilibrium with its melt and characterize the dynamics
of the SW. To do that we first generate an 
initial configuration, then simulate the system at coexistence and finally analyze the trajectories
generated in our simulations. Below we give some details about this procedure. 

\subsection{Generation of the initial configuration}

\begin{figure}
 \includegraphics[width=0.33\paperwidth,height=0.33\paperwidth,keepaspectratio]{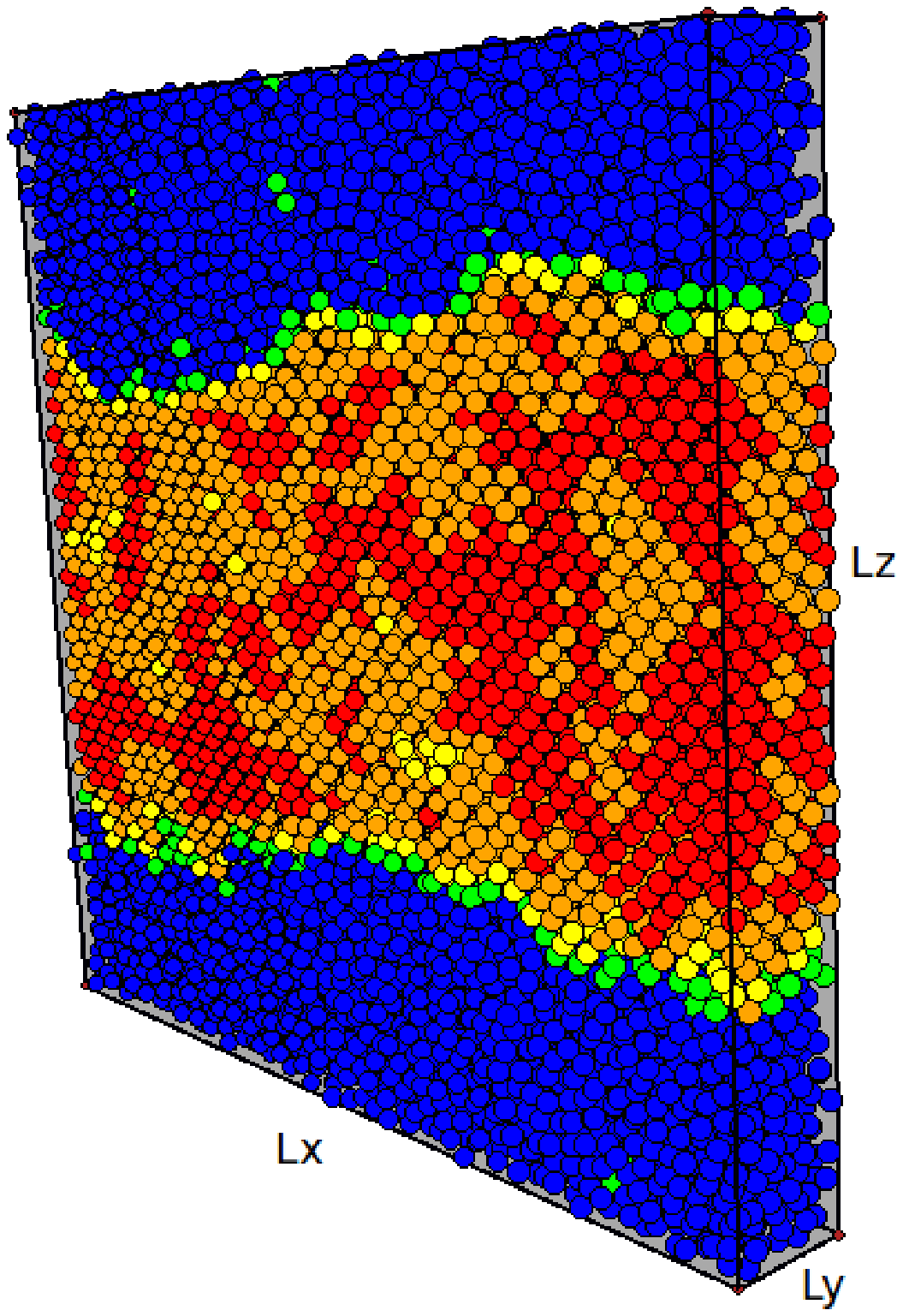} 
 \includegraphics[width=0.33\paperwidth,height=0.33\paperwidth,keepaspectratio]{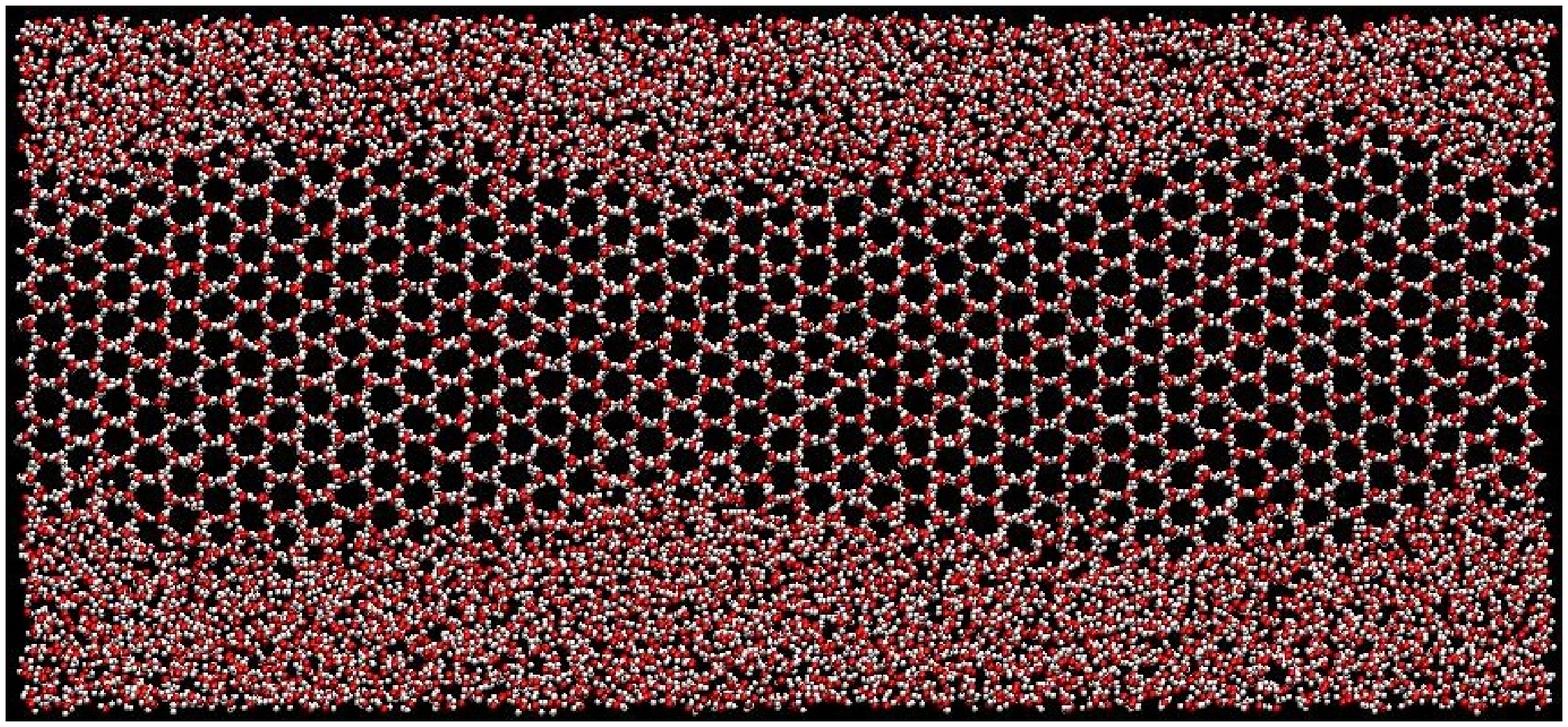} 
 \caption{Snapshots of the initial configuration for one of the LJ (top) and one of the water systems studied (bottom).
In the LJ system particles are coloured according to the extent to which their local environment resembles that of an
fcc lattice (in decreasing order of fcc-like character:
red, orange, yellow, green and blue).}
 \label{snapshot}
\end{figure}

The first step consists in creating an initial configuration.  Snapshots of an
initial configuration for LJ and water are given in Fig. \ref{snapshot}.  In
the snapshot corresponding to the LJ system we show the way we refer to the
edges of the simulation box, $L_x, L_y$ and $L_z$.  
By preparing systems as indicated in Fig. \ref{snapshot} we study the
dynamics of the CMI for the crystal face exposed in the $x$-$y$ plane
and SW propagating along the $x$ direction.
This sort of box geometry has already been used in a number of simulation
studies of the CMI (see, e. g., Refs. \cite{PhysRevLett.86.5530,Davidchack06,JCP_2003_119_03920,0295-5075-93-2-26006}).  
To specify the crystal orientation we indicate in parenthesis the 
Miller indexes of the interfacial plane ($x-y$), 
%parallel to the $x$-$y$ simulation box plane, 
and in square brackets the Miller indexes of the crystalline plane 
parallel to the direction of propagation and perpendicular to the interface
($x-z$).
%parallel to the $x$-$z$ simulation box plane. 
%
%e use the crystallographic frame to specify the crystal orientation. 
%or that purpose, we indicate in parenthesis 
% vector normal to the interface and in square brackets 
% vector normal to both the former vector and the 
%he direction of propagation of the waves. 

In Table
\ref{systems_size} we summarize the orientation and 
the size of the systems investigated. 

\begin{table}
 \footnotesize
  \begin{tabular}{c c c c}
  \hline
  \hline
  Model & Orientation & $L_x$x$L_y$x$L_z$ ($\sigma^3$) & Molecules \\
  \hline
  \multirow{2}{*}{HS}
     & (100)[001] & 47.046x4.705x47.046 & 10256 \\
     & (110)[001] & 44.359x4.705x52.058 & 10726 \\
  \hline
  \multirow{2}{*}{LJ}
     & (100)[001] & 49.101x6.336x49.181 & 14748 \\
     & (111)[11\={2}] & 50.524x6.7313x49.292 & 16160 \\
  \hline
  \multirow{4}{*}{TIP4P/2005}
	     & (basal)[prismI] & 59.418x5.710x29.541 & 10112 \\
	     & (prismI)[basal] & 57.024x6.962x25.581 & 10240 \\
	     & (prismII)[prismI] & 58.150x5.710x26.569 & 8896 \\
	     & (prismII)[basal] & 56.959x6.979x26.552 & 10670 \\
  \hline	     
  \hline
	     
 \end{tabular}
 \caption{Orientation and size of the systems investigated.}
 \label{systems_size}
 
\end{table}

To generate the initial configuration we first start by equilibrating the solid 
phase with  an $NpT$  simulation at the coexistence
pressure and temperature. For the equilibration of ice, ordinary Monte Carlo
moves were supplemented with full ring reorientation in order to properly
sample the hydrogen bond network \cite{rick03,macdowell10}. 
The final snapshot of the bulk solid thus equilibrated 
is rescaled so that the density matches the average equilibrium
density.  It is important to set  $L_x$ and $L_y$ equal to their respective
equilibrium values to avoid the solid being stressed in the $x$-$y$ plane
\cite{Frenkel_darkside}.  Then, a configuration of the fluid phase is
equilibrated in an $Np_zT$ simulation in a box with the same edges $L_x$ and
$L_y$ as the equilibrated solid. In the $Np_zT$ ensemble the pressure is 
exerted along the $z$ direction. 
In this way, the box keeps the sides $L_x$ and $L_y$ fixed so it can be
subsequently glued to the equilibrated solid along the $x$-$y$ plane.  The fluid 
is equilibrated at the coexistence temperature and at a pressure higher than
the coexistence pressure.  After equilibration of the fluid, we bring the solid
and the fluid boxes together and remove the fluid particles that are less than
a diameter apart from any solid particle.  This causes a small drop in the
fluid's density, which is partly compensated by the fact that the fluid was
equilibrated at a pressure higher than the coexistence pressure.  Finally, the
system is further equilibrated in the $Np_zT$ ensemble at the coexistence
temperature and pressure to ensure that the fluid's density fully relaxes to its coexistence
value.  The overall density of the initial configuration thus generated must
lie in between the densities at coexistence of the liquid and the solid (in
Table \ref{coexistence_values} we summarize the coexistence conditions for the
three models investigated).

\begin{table}
\footnotesize
\begin{tabular}{c c c c }
\hline
\hline
 Property & HS & LJ & TIP4P/2005 \\
\hline
 T & - & 1.0 $\epsilon/k_B$ & 252 K \\
 p & 11.54 $k_BT\sigma^{-3}$ & 4.95 $\epsilon\sigma^{-3}$ & 1 bar \\
 $\rho_c$ & 1.0369 $\sigma^{-3}$ & 1.005 $\sigma^{-3}$ & 0.921 $gcm^{-3}$  \\
 $\rho_l$ & 0.9375 $ \sigma^{-3}$ & 0.923 $\sigma^{-3}$ & 0.993 $gcm^{-3}$ \\
 \hline
 \hline
\end{tabular}
\caption{Coexistence values for the temperature, the pressure and the density  
for the different models.
The values for the HS, LJ, and TIP4P/2005 models were taken respectively from Refs. \cite{Noya08},
\cite{davidchack:7651} and \cite{C1CP22168J}. 
}
\label{coexistence_values}
\end{table}

\subsection{Simulation details}
\label{simdet}
Once we have an initial configuration, production runs are carried out in the
$NVT$ ensemble.  Given that the overall density of the system lies in between 
the coexistence densities of the two phases, the interface is stable in
an $NVT$ simulation at coexistence temperature. 

\textcolor{black}{In
principle, $NVE$ simulations provide correct trajectories that preserve local
momentum conservation and yield correct hydrodynamics. However, temperature
control on long $NVE$ simulations is difficult, particularly for systems exhibiting
two phase coexistence.  
\textcolor{black}{Of course, a sufficiently small time step
can always be chosen that will garantee numerical stability in a single, long $NVE$ run, 
but this would
result in prohibitively large CPU time. A possible strategy would be to obtain independent configurations 
of the crystal-melt
interface in a long $NVT$ simulation and use such configurations for 
short production runs in the $NVE$ ensemble.
However, we find  we can obtain correct results just from the long $NVT$
simulation (see Appendix \ref{nvt-nve}) by using a recently developed version of
the velocity-rescaling thermostat due to Bussi, Donadio and Parrinello \cite{bussi07}.
Such thermostat, perturbs the dynamics gently by effectively rescaling
the velocities over a large period and  has
met wide acceptance. 
}
Theoretical and numerical studies show
that this thermostat requires close to minimal perturbation of the correct time
evolution for given thermostating performance \cite{leimkuhler11}.  In
practice, Bussi et al. have shown that this algorithms provides diffusion
coefficients that are insensitive to the thermostating relaxation time chosen
in a range spaning several orders of magnitude.  Similarly, Delgado-Buscalioni
et al. note that a thermostat with a sufficiently large thermalising time
provides capillary wave dynamics of the liquid--vapor interface that does not
differ significantly from results performed in the $NVE$ ensemble
\cite{buscalioni_PRL_08}.  In order to be in the safe side, we employed
relatively large relaxation times of $\tau=1ps$ and $2ps$ for the Lennard-Jones and
water systems respectively. The relatively large system sizes that are required to perform
our study also help to achieve correct thermostatization with minimal
perturation of the dynamics \cite{bussi07,leimkuhler11}. Moreover, for the Lennard-Jones
system we have checked that changing the relaxation time from $\tau=1ps$ to
$\tau=100ps$ does not change our results.}

For the HS model, production runs were carried out using both a conventional
Monte Carlo (MC) algorithm and an event driven molecular dynamics (MD)
algorithm based on that provided in Ref. \cite{rapaport}.  In order to have
enough statistics we simulate $\sim$1000 trajectories starting from 
different initial configurations for MD simulations and $\sim$250 for MC.  Every
MD trajectory is run for $\sim4\e{6} (\sigma^2 m/k_BT)^{\frac{1}{2}}$, where
$\sigma$ is the particle diameter, $m$ the mass and $k_B$ the Boltzmann 
constant. 
For MC simulations we performed 1.5\textperiodcentered$10^6$ MC cycles where 
each cycle consists in an attempt of centre of mass displacement per particle. The 
maximum displacement for the centre of mass was set to 3.8\textperiodcentered$10^{-2}\sigma$.
In each trajectory (both MD and MC), 150 configurations were saved in a logarithmic time scale to
perform the subsequent analysis. 

To simulate water we used the MD GROMACS package \cite{GROMACS1,GROMACS2} and
the TIP4P/2005 water model \cite{abascal05}.  The time step for the
Velocity-Verlet integrator was fixed to 0.003 ps and snapshots were saved every
75 ps.  
Simulations were run for a total time of $\sim0.5 \mu s$. 
The temperature was set to 248.5 K. At this temperature, very close to 
the melting value of 252 K reported in Ref. \cite{abascal05}, we found no 
significant drift of the average height of the ice-water interface.

The LJ system was simulated using the MD GROMACS package.  We use
the truncated and shifted LJ potential proposed by Broughton and
Gilmer \cite{Broughton-lj}. We simulate the GROMACS implementation for Ar:
$\sigma = 3.405$ \r{A}, $\epsilon / k_B = 119.87$K, $m = 6.69\cdot10^{-26} kg$.  The time
step for the Velocity-Verlet integrator was fixed to 0.01 ps and snapshots were
saved every 2 ps for a total simulation time of 0.1 $\mu s$.  

For the case of HS we performed several independent short trajectories, whereas
for water and LJ we opted for running just one long simulation for each
investigated interface. 

In order to be able to compare the relaxation dynamics of the different simulated systems
in the same time scale we define the following dimensionless 
time: $t^* = t 6D/\sigma^2$, where $D$ is the diffusion coefficient of the fluid
at coexistence. The ratio $\sigma^2/(6D)$ is the average time it takes for a fluid
particle to diffuse its own diameter and we refer to it as 
``diffusive time''. Therefore, $t^*$
indicates  
the number of times a fluid particle diffuses its own diameter.  
\textcolor{black}{For the self diffusion coefficient of the fluid at coexistence we use 
$D=0.024(k_BT\sigma^2/m)^{1/2}$ for the HS model with event driven MD \cite{Davidchack98_HS}, 
$D=1.456\cdot10^{-5}~\sigma^2/$cycle for the HS model with MC simulations, 
$D=3.87\cdot10^{-3}~\sigma^2ps^{-1}$ for the LJ model (in good agreement with 
the value reported in Ref. \cite{Heyes88_lj}), and $D=0.3865~nm^2ns^{-1}$ 
for the TIP4P/2005 model.}

\subsection{Dynamics of the surface waves}

As shown in Fig. \ref{snapshot}, the CMI of the systems 
here investigated is wavy. The purpose of this work 
is to characterize the dynamics of such waves.
To do that we first define the local  interface 
position, or interface profile
$h(x_n)$, at discrete positions $x_n$ along the $x$ direction 
(see below for further details).

The interface profile is then Fourier transformed, and Fourier
modes $h_q$ defined as:
\begin{equation}
 h_q=\frac{1}{N}\overset{N}{\underset{n=1}{\sum}} h(x_n)e^{iqx_n}
 \label{Fourier_transform}
\end{equation}
where $N$ is 
the number of discretization points along the $L_x$ side of the simulation box, 
and each wave mode is associated with a reciprocal space vector, $q$, 
that can take values $q=2\pi k/L_x$, where $k$ is a positive natural number.
Small $q$ vectors correspond to wave modes with a large wave length and viceversa. 

The time-dependent autocorrelation function of $h_q$ is then given by:
  \begin{equation}
  f_q(t)=\frac{\left<h_q(0)h_q(t)^{*}\right>}{\left<h_q(0)h_q(0)^{*}\right>}. 
\label{acf}
 %f_q(t)=\frac{\left<\left(h_{q_{t=0}}-\overline{h}_q\right)\left(h_{q_{t=t}}-\overline{h}_q\right)\right>}{\left<\left(h_{q_{t=0}}-\overline{h}_q\right)\left(h_{q_{t=0}}-\overline{h}_q\right)\right>}
    \end{equation}
This function gives information about the way a capillary wave
mode relaxes. 
It depends not only on $q$, but also on the 
orientation of the crystal with respect to the fluid: $f_q(t) \equiv f(t,q,(h,k,l),[m,n,o])$. 
In this paper we analyze the $q$-dependence of $f_q(t)$ for three different models
and for several orientations.

%To obtain this function we employ an order parameter (see Appendix A) which allows us to identify the crystal-like atoms of the system. 
%The next step is to divide the interface in a number of bins, N, centred at $x_n$ along the x axis. This bins' width,  $\Delta_x$,  
%is an adjustable parameter which is determined for each system. 
%Then the location of the interface for each bin centred at $x_n$, 
%is calculated as the average height of the outer crystal-like atoms , averaged again over the short axis.
%Finally the values of these amplitudes are averaged for the two interfaces created due to the periodic boundary conditions.

\subsection{Determination of the interface profile, $h(x_n)$}
%In order to obtain the function $h(x_n)$ describing the interface location we divide each of the interfaces created due to 
%periodic boundary conditions in two strips of width $\Delta_y=L_y/2$ along the short axis. Then each one of these strips is divided 
%along the x axis in a number of bins N, centred at $x_n$, of width $\Delta_x$, which is an adjustable parameter. 
%The interface location at each point $x_n$ is then calculated as the average height of the outer atoms of each grid, 
%averaged again along the short direction. 
\textcolor{black}{The definition of a suitable interface profile from a set of atomic positions is a subttle matter \cite{chacon09,miguel10}.
It is now well understood that the evaluation of the function $h(x_n)$ consistent with the capillary wave model
requires to properly identify the phase to which atoms may be
attributed, and only then, searching for an optimal surface separating each phase \cite{chacon09,miguel10}. Whereas the optimal
process is involving and time consuming \cite{miguel10}, it has been observed that dynamic properties are rather insensitive
to details of the specific procedure \cite{buscalioni_PRL_08}. For this reason, we have chosen a simple method, inspired on that 
proposed in Ref. \cite{Davidchack06}, that
is computationally convenient and is
very robust to the arbirary parameters  required in practice (c.f. section \ref{analparam}).}

To obtain the discrete function $h(x_n)$ describing the profile of the interface
along the $x$ direction we consider the outermost particles of the crystal slab. 
We first label the molecules in the system as fluid-like or solid-like. 
To do that we make use of local bond order parameters 
that are able to distinguish between fluid and solid-like
particles in an instantaneous configuration by looking
at the relative position of a particle with respect to its neighbours
(see Appendix \ref{ordparam} for details).

%%%%%%%%%%%%%%%%%%%%%%%%%%%%%%%%%%%%%%%%%%%%%%%%%%%%%
% alternativa
%%%%%%%%%%%%%%%%%%%%%%%%%%%%%%%%%%%%%%%%%%%%%%%%%%%%%
%To do that we make use of local bond order parameters 
%that are able to distinguish between fluid and solid-like
%particles by looking at the relative position of a 
%particle with respect to its neighbours(see Appendix A for details). 
%This is done in every instantaneous configuration.
%%%%%%%%%%%%%%%%%%%%%%%%%%%%%%%%%%%%%%%%%%%%%%%%%%%%%%%

Once all molecules are labelled, we remove the fluid-like particles and
among all solid-like particles 
(red, orange, and yellow
in Fig. \ref{snapshot} (top))
we take those that form the largest cluster. 
In this way we are left with the crystal phase alone.  
Note from Fig. \ref{snapshot} that due to the geometry of the system we have two 
independent interfaces. We explain below how we calculate $h(x_n)$ for one of them.

We start by splitting $L_y$ in two, so the interface
is divided in two elongated stripes. 
Each stripe is divided in $N$ equispaced points along 
$L_x$. 
These points define the set of $x_n$ values in which $h(x_n)$ is evaluated. 
For a point with coordinates $(x_n,y_p)$, with $p={1,2}$ indicating a given
stripe,
the local amplitude $h(x_n,y_p)$ is evaluated by averaging the 
$z$ coordinate of the $n_o$ 
outermost atoms with $y$ coordinate $\in [y_p-L_y/4:y_p+L_y/4]$ and $x$ coordinate
$\in [x_n-\Delta_x/2:x_n+\Delta_x/2]$ ($\Delta_x$ and $n_o$ 
are adjustable parameters). In this way, a function $h(x_n,y_p)$ is obtained for each stripe, and 
the final $h(x_n)$ is obtained as the average between the stripes corresponding to $y_1$ and $y_2$.

Thus, the adjustable parameters to obtain a discretized profile of the interface in
the way above described are $N$, $\Delta x$, and $n_o$. 
The results shown in the remaining of the paper correspond to the following set
of analysis parameters: $N=50$ , $\Delta x=3\sigma$ and $n_o=4$. 
In section \ref{analparam} we show that our main results are not affected
by this particular choice of analysis parameters. 

\subsection{Interfacial stiffness}

To test our simulations we make use of the 
following expression provided by Capillary Wave Theory
\cite{jasnow84,nelson04}:
\begin{equation}
 \left<|h_q|^2\right>=\frac{k_BT}{A\widetilde \gamma q^2}
\label{eqstiffness}
\end{equation}
that  relates the average squared amplitude of the capillary wave mode $q$, $\left<|h_q|^2\right>$,
to the interfacial stiffness, $\widetilde \gamma$, 
by means of the equipartition
theorem (note from Eq. \ref{acf} that $\left<|h_q|^2\right>$ is equal to 
the unnormalized $f_q(0)$). In the equation above $k_B$ is the Boltzmann constant
and $A=L_x L_y$ is the area of the interface. 
The interfacial stiffness has been carefully obtained for a number of
systems \cite{Davidchack06,0295-5075-93-2-26006,jcp_aleksandar_yukawa,Oettel12}, with the HS model among them. Therefore, we can
double-check our results by comparing our value for $\widetilde \gamma$ with that obtained
in Refs. \cite{Davidchack06,Oettel12}.

\subsection{Kinetic coefficient}
\label{kincoef}

An important parameter in crystal growth is the kinetic coefficient, $\mu$. 
The kinetic coefficient of a CMI is the proportionality constant
between the speed at which the interface front advances, $v$, and the supercooling, $\Delta T$: 
\begin{equation}
v = \mu \Delta T
\end{equation}
where $\Delta T = T_m-T$ is the 
the difference between the melting temperature, $T_m$, and the temperature of interest, $T$.

As shown in Refs. \cite{karma93,hoyt02}, by analysing the 
crystal-melt CW, it 
is possible to obtain an estimate of $\mu$. The method entails first obtaining
$f_q(t)$ via Eq. \ref{acf} for a number of $q$-modes, then fitting each $f_q(t)$
to an exponential function of the type $\exp(-t/\tau_q)$ to get a characteristic
decay time $\tau_q$ for each mode, and finally obtaining
$\mu$ from the slope of a representation of $1/\tau_q$ vs $q^2$:
\begin{equation} 
1/\tau_q = \frac{\mu \widetilde \gamma T_m}{\Delta h_m \rho}  q^2,
\label{mu} 
\end{equation} 
where $\Delta h_m$ is the molar melting enthalpy, $\rho$ is the crystal density, 
and $\widetilde \gamma$ is the stiffness, that can be obtained via Eq. \ref{eqstiffness}
by extrapolating $\widetilde \gamma (q)$ to $q=0$.  

Following this method, in this work we compute the kinetic coefficient for all
the interfaces investigated. We compare our results for HS with those obtained
in Ref. \cite{amini06} using the same technique, and our results for water with
those obtained in a recent publication using a different approach
\cite{kusalik2011}. 

\section{Results and discussion}

\subsection{Dynamics of crystal--melt surface waves}

\begin{figure}
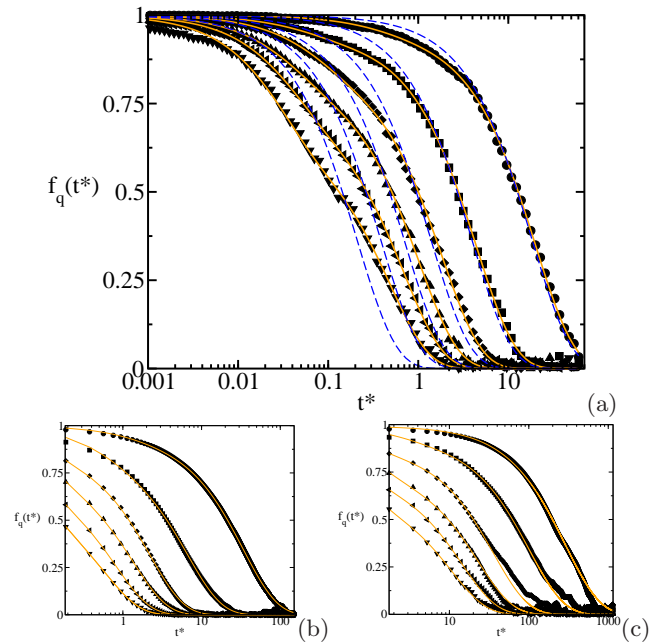

\includegraphics[width=0.40\textwidth,keepaspectratio,clip]{ajuste.dobleexp.hs.100.md.rescaled.eps}(a)
\includegraphics[width=0.21\textwidth,keepaspectratio,clip]{ajuste.dobleexp.lj.100.rescaled.eps}(b)
\includegraphics[width=0.21\textwidth,keepaspectratio,clip]{ajuste.dobleexp.pII.basal.rescaled.eps}(c)
\caption{(a)-(c), symbols: autocorrelation functions for the HS (100)[001], 
the LJ (100)[001], and the water (pII)[basal] interfaces respectively (all data correspond to MD simulations).  
In a given plot, curves from right to left correspond to wavevectors
$q=2\pi k/L_x$, with $k \le 6$. Lines correspond to different fits: dashed blue to a single exponential, 
and  solid orange to a double exponential.}
\label{ajustes}
\end{figure}

%%%%%%%%%%%%%%%%%%%%%%%%%%%%%%%%%%%%%%%%%%%%%%%%%%%%%%%%%%%%%
% VERSION 11
%%%%%%%%%%%%%%%%%%%%%%%%%%%%%%%%%%%%%%%%%%%%%%%%%%%%%%%%%%%%%
For each system described in Table \ref{systems_size} we evaluate  $f_q(t)$
(Eq. \ref{acf}) for several values of $q$. Some of these autocorrelation
functions are shown in  Fig. \ref{ajustes} (a), (b) and (c) for a HS, an LJ and a water
CMI respectively. 
As expected, the correlation functions decay from 1 to 0 as the wavemodes
relax, and the relaxation for a given interface is the slower the
smaller the wave vector $q$ (or the larger the wavelength). It is also apparent
that for the wavelengths studied, 
the surface wave dynamics corresponds to a strongly damped regime, with no signs
of oscillatory behaviour in any of our the autocorrelation functions. 

This is consistent with previous simulation studies of the CMI, 
where the decay of the
correlation functions were found to be purely exponential, or at least showed a
purely monotonous decay \cite{Hoyt20101382,amini06}. Taking this into account,
we first attempt to describe the correlation functions by a single
pure exponential $f_q(t)=e^{-t/\tau_{exp}}$, 
where the characteristic decay time 
$\tau_{exp}$ is the only fitting parameter. However, as seen in Fig.
\ref{ajustes} (a) (dashed blue curves), an exponential fit does not accurately
describe the decay of $f_q(t)$, particularly for  curves corresponding to 
large $q$s.  

Considering that SW could exhibit very different behaviour at high and
low frequencies, we then attempted to fit our results using a double
exponential:
\begin{equation}
f_q(t)=Ae^{-t/\tau_{ds}}+(1-A)e^{-t/\tau_{df}}, 
\label{de}
\end{equation}
where $A$, 
$\tau_{ds}$ (characteristic time for a slow relaxation process) and
$\tau_{df}$ (characteristic time for a fast relaxation process)
are the fitting parameters.  As
it can be seen in Fig. \ref{ajustes} (a) (orange solid lines) this fit accurately describes
all curves and is significantly better than a single 
exponential (we also tried fitting our results to a stretched exponential, but the
resulting fit was not as good as that of a double exponential and is not shown). 
The double exponential fit does a good job for all systems investigated. 
To illustrate this, in Figs. \ref{ajustes}(b) and (c) we show the correlation functions alongside
their corresponding double exponential fits for the LJ (100)[001] and the water (pII)[basal] 
interfaces respectively. For the case of water the simulations are slower than for the other systems 
and gathering statistics
to obtain quality data for $f_q(t)$ at long times is a very involving task.  
As a consequence, the values of $f_q(t)$ for the ice-water interface at long times are rather noisy 
and have not been taken into account to obtain the fits shown in Fig. \ref{ajustes}(c). 

The adequacy of the double exponential fit suggests the existence of two distinct relaxation
time scales: a fast one responsible for the initial decay and a slow one responsible 
for the decay at long times. 
The presence of two
simultaneous relaxation
time scales resembles the behaviour observed at the interface of viscoelastic
materials \cite{tejero85,pleiner88,harden91,cao91,dorshow93,monroy98}, where
a high frequency relaxation process is related to 
elastic Rayleigh waves, while that of low frequency is related to CW. 
Indeed, it has been shown that both elastic Rayleigh waves and capillary
Kelvin waves may exhibit an overdamped
regime where oscillations are completely suppressed and the relaxation is 
exponential \cite{tejero85}. By analogy, we assume in principle that the two 
different time scales found in our study for the CMI are associated
to different relaxation mechanisms. 
 
\begin{figure}
\includegraphics[width=0.40\textwidth,keepaspectratio,clip]{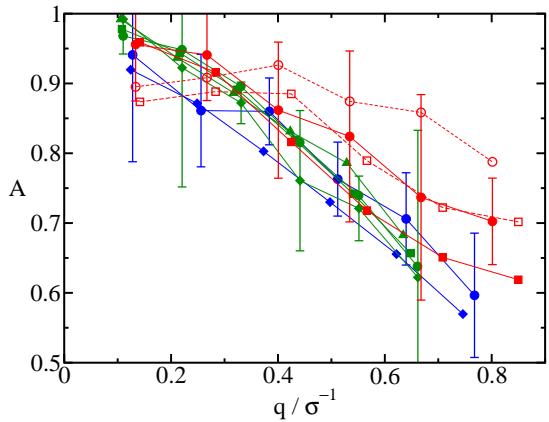}
\caption{Prefactor for the double exponential fit (see main text) as a function 
of $q$ for all systems investigated. HS, LJ and water data are shown in red, blue and green respectively. 
HS data are shown for the (100)[001] (circles) and
(110)[001] (squares) interfaces; LJ 
for the (100)[001] (circles) and (111)[11\={2}] (diamonds) interfaces;
and water for the 
(basal)[prismatic I] (triangles), (prismatic I)[Basal] (circles), (prismatic II)[prismatic I] (squares)
and (prismatic II)[basal] (diamonds) interfaces. 
For the HS system filled (empty) symbols and solid (dashed) lines correspond
to MD (MC) simulations. \textcolor{black}{Error bars are shown for one of the orientations
of each model potential. Different orientations of the same model have similar error bars but have
not been shown for clarity.}} 
\label{coefa0}
\end{figure}

The parameter $A \in [0:1]$ in Eq. \ref{de} quantifies the weight of each mechanism in the relaxation of
CMI waves.  
When $A$ is close to 1 the decay of $f_q(t)$ is dominated by the slow process and when
it approaches 0.5 the decay of $f_q(t)$ is affected by both slow and fast processes. 
In Fig. \ref{coefa0} we plot $A$ as a function of $q$ for all systems investigated.
In all cases $A$ is close to 1 for the smallest $q$ investigated 
and decreases as $q$ increases. 
Therefore, we observe a relaxation essentially dominated by the slow process at low $q$ (large wavelengths) 
and affected by both slow and fast processes at large $q$ (small wavelengths).

We first attempt to elucidate the nature of the slow relaxation process
by analysing the dependence of $\tau_{ds}$ on $q$. 
By carefully fitting our autocorrelation functions $f_q(t)$ to Eq. \ref{de} 
we obtain $\tau_{ds}(q)$  (Note that obtaining meaningful parameters
from a double exponential fit is not trivial. We had to address this issue
carefully and give some indications of the fitting  procedure in Appendix 
\ref{defit} ). 
In Fig. \ref{half-life time} (a) we represent $\tau^*_{ds}$ versus $q$  for all interfaces investigated. 
In a double logarithmic scale it appears that all curves are parallel to each other within the
accuracy of our calculations. 
This suggest the existence of a power law of the type $\tau_{ds} \propto q^\alpha$, with 
the $\alpha$ exponent common to all systems. 
A power law is an indication that there is a common
mechanism underlying the slow relaxation process of all interfaces investigated. 
This is remarkable taking into 
account the different nature of the systems here studied. 
A visual inspection of  the relaxation of large wavelength SW, 
those more clearly affected by the slow
process, 
suggests that such process can be identified with 
the continuous recrystallization and melting taking place at the interface 
(see  online movie \cite{movie}).
 
In 1993 Karma published a theory for the relaxation dynamics of crystal-melt
CW based on a diffusion equation of the interfacial profile \cite{karma93}.
Karma's 
theory predicts a power law relation between a characteristic relaxation time and $q$. The 
obtained theoretical value for the exponent is $\alpha = -2$. In Fig. \ref{half-life time} (a) we include a dashed line
with slope -2 in  the double logarithmic representation. Within the accuracy of our calculations
all curves look parallel to the dashed line. Therefore, the dynamics of the slow process 
is consistent with Karma's theory \cite{karma93}. 
This implies that we can identify the slow process with the relaxation of CW. 
In other words, the slowly relaxing SW are in fact CW. 
The description of the relaxation 
of CW via a diffusion equation in Ref. \cite{karma93} 
is consistent both with the view inferred from our movies
that the slow process is due to the recrystallization/melting at the interface and 
with the absence of oscillations in our $f_q(t)$s.

The theory of Karma
has been previously tested in simulations of metals \cite{hoyt02,hoyt03,Hoyt20101382,monk2010} and 
HS \cite{amini06}. 
In these works it
was suggested that the relaxation of crystal-melt SW 
can be approximated by a single exponential for all $q$s. Our results 
above show that the scenario is more complex due to presence of 
fast relaxation processes that clearly affect    
large $q$ modes. 

\begin{figure}
\includegraphics[width=0.33\paperwidth,height=0.33\paperwidth,keepaspectratio,clip]{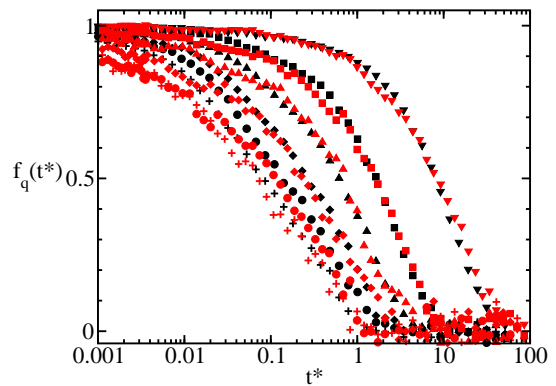}
\caption{Autocorrelation functions for the (110)[001] HS interface obtained by MD
(black) and MC (red) simulations. From right to left functions
corresponding to wavevectors $q=2\pi k/L_x$, with $k \le 6$,
are represented.} \label{correl-MC_vs_MD} \end{figure}

We can gain further insight into the nature of the slow process by
studying the relaxation of the HS CMI with MC simulations. 
In Fig.\ref{half-life time} (a) we compare the
slow relaxation times, in diffusive units, of HS as obtained by MD and MC simulations. 
The data coming from both simulation techniques lie on top 
of each other. The agreement between both simulation techniques is
further confirmed in Fig. \ref{correl-MC_vs_MD}, where we compare
the whole autocorrelation function for several $q$s.
The superimposition between MC and MD curves 
in diffusive time units points out the relevance
of the diffusive time scale for the relaxation of CMI SW. 
Moreover, 
\textcolor{black}{we can tell from the good agreement between MC and MD that the microscopic dynamics
is not playing any significant role in the relaxation of CMI CW}. 
This finding strongly contrasts with the case of the fluid-fluid
interface \cite{jeng98, buscalioni_PRL_08}, although it not so surprising if one takes into account that the crystal  
has an infinitely large viscosity.  

\textcolor{black}{As mentioned in the introduction, we are not aware of a hydrodynamic theory of
CW for the CMI. For lack of a better theoretical framework, we discuss here
our observations in the context of the CW theory for either a dense or a
viscoelastic medium with air \cite{landau91,jeng98,tejero85}.
Such systems present first a weakly damped regime at low $q$, with a damped
oscillatory behavior of frequency $\omega\propto q^{3/2}+iq^2$ (Kelvin waves).
At larger wavelengths, there is a crossover to a strong damping regime, where
the oscillations are completely supressed and the frequency becomes purely
imaginary  $\omega\propto iq$. In our simulations, we observe only a strongly
damped regime, but, at odds with standard theories for the liquid--vapor interface,
the damping is not linear in $q$, but rather, decays as $q^2$. Our observation
does not rule out the possibility of the aforementioned regimes occurring at
smaller wavevectors than are accesible to our simulations. However, it is clear that
the damping regime we observe is not the standard overdamped regime linear in $q$ that
has been reported elsewhere for the liquid-vapor interface \cite{buscalioni_PRL_08}.  Also note that a recent experimental
study of the CMI of strongly repulsive colloids  reported  relaxation dynamics in agreement
with a  linear $q$ dependence \cite{Weeks09}, but  neither our
results nor those of Ref.\cite{amini06} seem to support this conclusion.
Interestingly, Jeng et al.
noticed in their hydrodynamic theory of surface waves a strongly damped regime with
purely imaginary frequency which has both a quadratic and linear contribution in $q$.
However, we do not observe any signature of mixed quadratic and linear dependence
in our relaxation times either. Possibly, only a hydrodynamic theory incorporating both
the viscous behavior of the liquid and a viscoelastic response of the crystal can
predict the behavior observed in our simulations \cite{tejero85}.}

Fig. \ref{half-life time} (b) displays the characteristic times for the fast
relaxation process, $\tau^*_{df}$, as a function of the wave vector.  
For a given
$q$, $\tau^*_{df}$ is one or two orders of magnitude lower than $\tau^*_{ds}$.
Both $\tau^*_{df}$ and $\tau^*_{ds}$ decrease as $q$ increases.  
\textcolor{black}{Altghouh the error bars for $\tau^*_{df}$ are quite large, it seems that 
a power law is not evident from Fig.
\ref{half-life time} (b).}  This suggests that the fast relaxation can not be
identified with a single process but rather with a combination of different
ones.  Such processes must be fast as compared to the diffusion of the
interfacial front ($\tau^*_{df} << \tau^*_{ds}$) and must cause small perturbations
in the interfacial profile (their effect vanishes as $q$ goes to 0, as shown in
Fig. \ref{coefa0}).  Having these characteristics in mind, processes such as 
Rayleigh SW,
subdiffusion of the fluid, or the attachment/detachment of single particles
to/from the crystal phase are likely to lie behind the fast relaxation of the interface. 

In summary, the relaxation of SW is best described 
by a double exponential in the $q$-range analyzed in this paper. 
This is indicative of the existence of two distinct relaxation time scales. 
The slow one corresponds to the overdamped relaxation of CW by diffusion 
of the interfacial front (the counterpart of overdamped CW for the fluid-fluid
interface, although with a different $\tau(q)$ dependence). 
The fast one is due to quick, small perturbations of the interface profile 
possibly caused by 
Rayleigh waves, subdiffusion, and 
the attachment/detachment of particles to/from the crystal. 
In the limit of $q=0$ the decay of $f_q(t)$ can be entirely described 
by a single exponential function corresponding to the slow relaxation process. 

%\begin{figure}
%\includegraphics[width=0.33\textwidth,keepaspectratio,clip]{../figuras/ajustes.eps}(a)
%\includegraphics[width=0.33\textwidth,keepaspectratio,clip]{../figuras/stretching_coeficients.eps}(b)
%\caption{(a) Fits for the wavevectors q=0.267$\sigma$ (circles) and q=0.534$\sigma$ (squares) for the (100)[001] HS system 
%obtained by MD simulations. Dotted and dashed blue line are fits to $f_q(t)=e^{-t/\tau_e}$, 
%dashed green line to $f_q(t)=e^{-(t/\tau_s)^\beta}$ and solid black line to 
%$f_q(t)=Ae^{-t/\tau_{ds}}+(1-A)e^{-t/\tau_{dl}}$. Inset: Representation of f$_q$(t) versus time for wavevector q=2$\pi/L_x$ for different systems ( from left to right HS (100)[001], LJ (100)[001] and
%TIP4P/2005 (pII)[Basal] ). We obtain a characteristic time, $\tau_{dl}$, from
%the slope of the straight region for each wavevector. Time is given in
%$(\sigma^2 m/k_BT)^{\frac{1}{2}}$ units for HS and in ps for LJ and TIP4P/2005
%models. TIP4P/2005 results have been rescaled to collapse in the same scale.
%(b) Stretching coefficients $\beta$ for all the systems and orientations studied. Legend is 
%the same as in Fig. \ref{half-life time}.}
%\label{ajustes}
%\end{figure} 

\begin{figure}
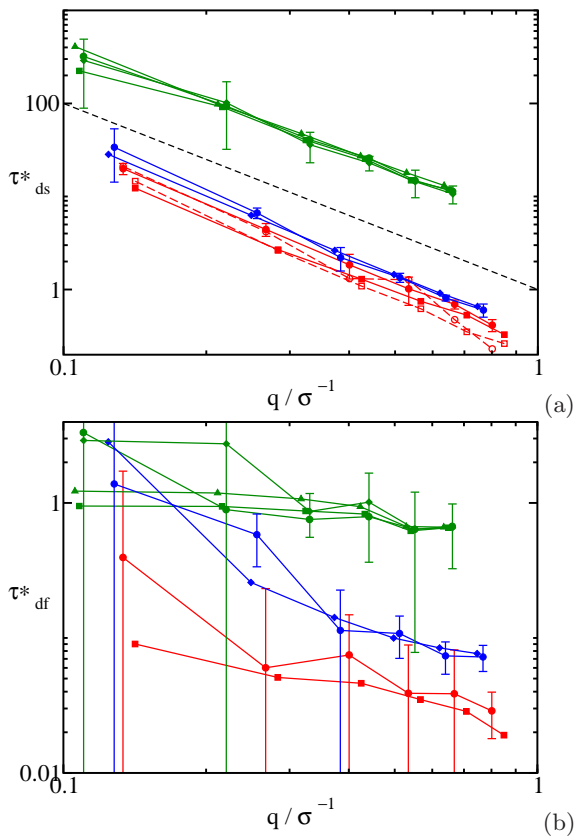

\includegraphics[width=0.33\paperwidth,clip]{tlargo_vs_q-articulo.rescaled.errores.eps}(a)
\includegraphics[width=0.33\paperwidth,clip]{tcorto_vs_q-articulo.rescaled.errores.eps}(b)
\caption{(a) Dimensionless characteristic time for the slow relaxation process, $\tau^*_{ds}$, plotted 
against the wave vector $q$ in a double logarithmic scale for all interfaces investigated.
Color code same as in Fig. \ref{coefa0}. (b) Same as (a) but for the fast relaxation process. 
The dashed line, included in (a) for visual reference, 
corresponds to a power law of the type $\tau_{ds} \propto q^{-2}$.} 
\label{half-life time} 
\end{figure}

We conclude this section by comparing the slow relaxation of different systems. Such comparison
is enabled by the use of a common time unit: the diffusive time.  
It is evident from figure \ref{half-life time} (a) that the curves corresponding to water
lie about an order of magnitude above those corresponding to LJ or HS. 
This implies that, for a given $q$, the water interface requires roughly ten times as much
diffusive time units
as LJ or HS particles in order for the interface to relax. For instance, for the 
smallest studied $q$ ice-water CW take to relax  the time
needed for a molecule to diffuse about 300 times its own diameter, while it only takes
about 30 times for the LJ or the HS systems. 
This difference in time scales is most likely related to the fact that 
water molecules, contrary to the case of LJ or HS particles, have orientational
degrees of freedom and need to be properly oriented to accommodate into the
crystal phase. 
%In any case, despite this time-scale difference between 
%systems with or without orientational degrees of freedom, 
%the power law $\tau_{ds} \propto \alpha^{-2}$
%holds in all cases and is governed by the diffusion of the fluid phase leading to
%the diffusion of the interface front.  

\subsection{Kinetic coefficient}

\begin{figure}
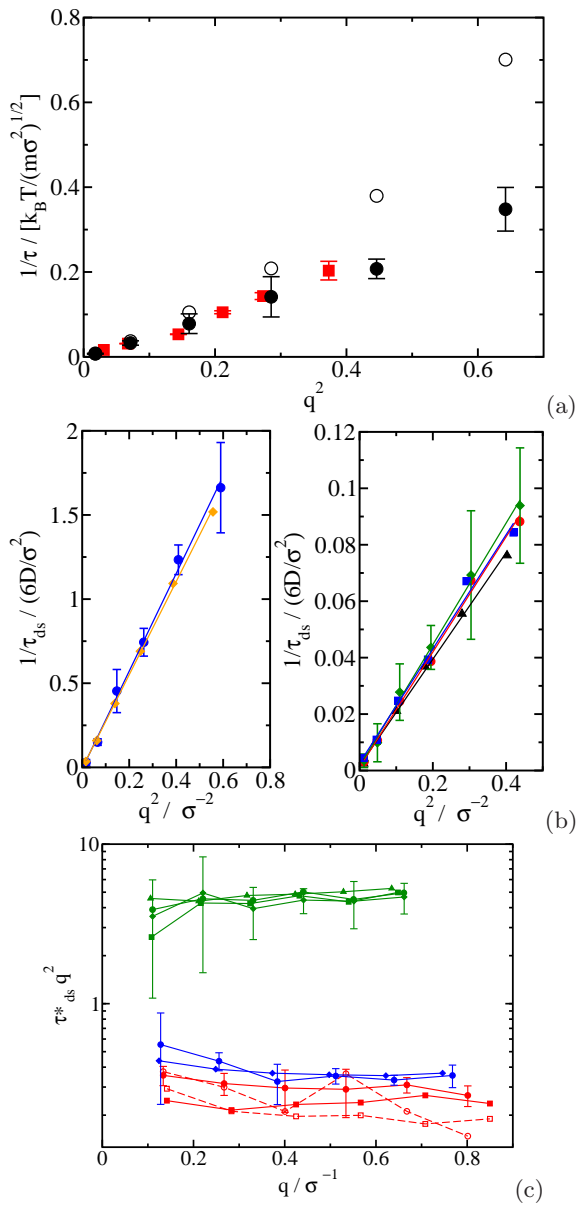

\includegraphics[width=0.33\paperwidth,clip]{comp_amini.errores.eps}(a)
\includegraphics[width=0.33\paperwidth,clip]{tau-1_vs_q2-rescaled_LJ-water.errores.eps}(b)
\includegraphics[width=0.35\textwidth,keepaspectratio,clip]{tlargoq2_vs_q-articulo.rescaled.errores.eps}(c)
\caption{Plots of the inverse relaxation time versus $q^2$. From the slope 
of these plots the kinetic coefficient can be inferred via Eq. \ref{mu} \cite{karma93,hoyt02}. In (a) 
our results for the (100)[001] HS interface (circles) are compared to those reported in 
Ref. \cite{amini06} (squares). Empty (solid) circles correspond to relaxation times
estimated by fitting $f_q(t)$ to a single (double) exponential. In (b) we 
show our results for all interfaces investigated for the LJ (left) and the water (right) systems respectively. 
\textcolor{black}{(c): $\tau^*_{ds}q^2$ as a function of $q$ in a
logarithmic scale for all interfaces investigated.}
Color code same as
in Fig. \ref{coefa0}}
\label{compamini}
\end{figure}

As explained in Sec. \ref{kincoef}, the kinetic coefficient, $\mu$, can be 
estimated by measuring the autocorrelation function $f_q(t)$ for a few CW modes 
 \cite{karma93,hoyt02}. Using this methodology Amini et al. 
estimated $\mu$ for the HS system \cite{amini06}. 
In Fig. \ref{compamini} (a) and (b) we plot the inverse of the relaxation time 
versus $q$ for an interface of the HS system and compare our results (filled circles)
with those of Ref. \cite{amini06} (filled squares). 
From the slope of the plot shown in Fig. \ref{compamini} one can obtain $\mu$ via
Eq. \ref{mu}. \textcolor{black}{(Alternatively, as shown in Fig. \ref{compamini} (c), a representation of $\tau_{ds}q^2$ vs $q$ gives a horizontal line
from which $\mu$ can be obtained)}. As shown in Fig. \ref{compamini}, the agreement with Ref. \cite{amini06} is quite satisfactory, 
which gives us great confidence in our calculations. The filled circles in Fig. \ref{compamini} 
were obtained by fitting our $f_q(t)$s to Eq. \ref{de} in order to get the characteristic time $\tau_{ds}$. 
If we get the characteristic relaxation time by fitting $f_q(t)$ to a simple exponential function,
disregarding the fact that there are two distinct time scales involved in the
relaxation of crystal-melt SW, 
we get the empty circles in Fig. \ref{compamini}, that are not in good agreement 
with Ref. \cite{amini06}.  Although, to the best of our knowledge, the need of
fitting $f_q(t)$ to a double rather than to a single exponential is pointed 
out for the first time in this work, previous studies that assumed a single exponential
behaviour provide results that are consistent with ours \cite{amini06}. This apparent 
contradiction is explained by the fact that in Ref. \cite{amini06}  
the characteristic time was obtained
from the slope of the linear regime in a plot of $\ln f_q(t)$ vs. $t$ \cite{amini06}, which, for long times, gives the 
characteristic time for the slow relaxation process 
(note that eq. \ref{de} can be approximated by a single exponential
for long times given that $\tau_{ds} >> \tau_{df}$). 
Such decoupling between the fast and the slow process is less and less
evident as the relaxation of the $q$ modes becomes faster or, equivalently, 
as $q$ increases. Therefore, in order to consider large $q$s 
for the calculation of $\mu$ from a plot like that shown in Fig. \ref{compamini}
it is advisable to use Eq. \ref{de} to obtain the characteristic time
for the slow relaxation process. In fact, with our analysis we are able
to extend the linear regime of $1/\tau_{ds}$ vs $q^2$ to larger $q$s than
in Ref. \cite{amini06}.   

In Fig. \ref{compamini} (b) we show a $1/\tau_{ds}$ vs. $q^2$ representation for the
LJ (left) and water (right) systems. All interfaces show, within the accuracy
of our measurements, a linear dependency of $1/\tau_{ds}$ vs. $q^2$.  This result
had already been anticipated in Fig. \ref{half-life time} (all data sets are
parallel to the dashed line).  From the slopes in Fig. \ref{compamini} we estimate the kinetic
coefficients, which we report in Table \ref{kincoeff}.  Since we reduce time by
the diffusive time (see Sec.  \ref{simdet}) and distance by the particle
diameter, the kinetic coefficient in our reduced units tells us how faster the
interface advances, in terms of diameters per diffusive time, when the
temperature is lowered by 1 $K$. The data in Table \ref{kincoeff} reveal that
the kinetic coefficient of water is more than an order of magnitude lower than
that of LJ. This means that one would need to supercool water ten times more
than LJ to get the same speed-up of the interface advance measured in
diameters per diffusive time. Our work shows that  
both the relaxation and the growth of the interface
are dramatically slowed down for the case of water, probably due to 
the role of orientational degrees of freedom in crystallization. 

 \begin{table}
 \footnotesize
  \begin{tabular}{c c c}
  \hline
  \hline
  Model & Orientation  & $\mu$/[$6D/(\sigma K)]$ \\
  \hline
    \multirow{2}{*}{LJ}
     & 100 & $(8\pm2)$\textperiodcentered$10^{-2}$ \\
     & 111 & $(5.8\pm0.6)$\textperiodcentered$10^{-2}$ \\
  \hline
  \multirow{3}{*}{TIP4P/2005}
	     & basal & $(2.1\pm0.3)$\textperiodcentered$10^{-3}$\\
	     & prismatic I & $(3.0\pm0.4)$\textperiodcentered$10^{-3}$ \\
	     & prismatic II & $(2.5\pm0.5)$\textperiodcentered$10^{-3}$ \\
  \hline  
  \hline
	     
 \end{tabular}
 \caption{Kinetic coefficient, $\mu$, for the LJ and TIP4P/2005 systems.}
 \label{kincoeff}
 
\end{table}

\begin{figure}
\includegraphics[width=0.33\paperwidth,clip]{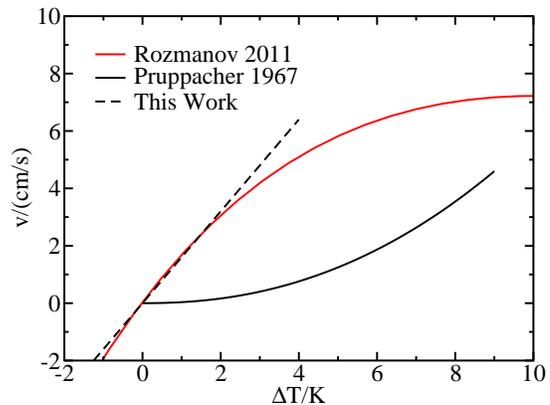}
\caption{Rate of crystal growth versus supercooling. We compare our 
results (dashed line) with the simulation results of Rozmanov et al. \cite{kusalik2011} (red solid curve) and 
with experimental data by Pruppacher \cite{pruppacher1967} (black solid curve).}
\label{growthrate}
\end{figure}

We can also compare our results for water with experimental \cite{pruppacher1967}
and simulation \cite{kusalik2011,berne2014} measurements of the speed of crystal
growth, $v$.  Using Eq. \ref{mu} and knowing that $v = 0$ for $\Delta T = 0$ we
can estimate $v$ {\it in the vicinity} of the melting temperature from our
calculated $\mu$.  In Fig. \ref{growthrate} we plot $v$ versus the supercooling,
$\Delta T$, for the basal plane of water, for which we obtained a kinetic
coefficient $\mu = 1.6~cm/(sK)$.  Our results correspond to the dashed line in Fig. \ref{compamini}. A
negative $v$ means that the interface recedes because the crystal melts for
$\Delta T < 0$.  The red solid curve corresponds to the simulation results reported in
Ref. \cite{kusalik2011} (large system), where $v(T)$ for the basal plane was
measured by monitoring the height of the interface
along time for different temperatures. Remarkably, our estimate of $v$ around
the melting temperature is in very good agreement with the results of Ref.
\cite{kusalik2011},  obtained by a completely different approach.  It should be
noted that the proportionality law of Eq. \ref{mu} only works for a narrow
temperature range around melting. Outside that range, the dependence of $v$
with $\Delta T$ is not linear any more and even shows a maximum at $\Delta T \approx 12~K$
\cite{kusalik2011}.  Our kinetic coefficients for the prismatic I, and prismatic
II planes are $\mu=2.2~cm/(sK)$ and $\mu=1.8~cm/(sK)$ respectively. The
latter is consistent with a recent simulation study
of the speed of crystal growth for the prismatic II plane \cite{berne2014}.  The comparison of the model
with the experiment \cite{pruppacher1967}, solid black curve in Fig.
\ref{growthrate}, is not as satisfactory, though.  The experimental $v$ near
coexistence is much lower than that predicted by the model.  
Further work is needed to understand why 
a model that gives a good agreement with the experiment for
the rate of crystal nucleation \cite{jacs2013} is not able to accurately predict the
rate of crystal growth. 

\subsection{Robustness of our calculations}

In order to check if our calculations are robust we asses the 
dependence of our results on both the choice of the analysis parameters and 
the system size. Moreover, whenever possible, we compare our results with the 
existing literature. 

\subsubsection{Analysis parameters}
\label{analparam}

\begin{figure}
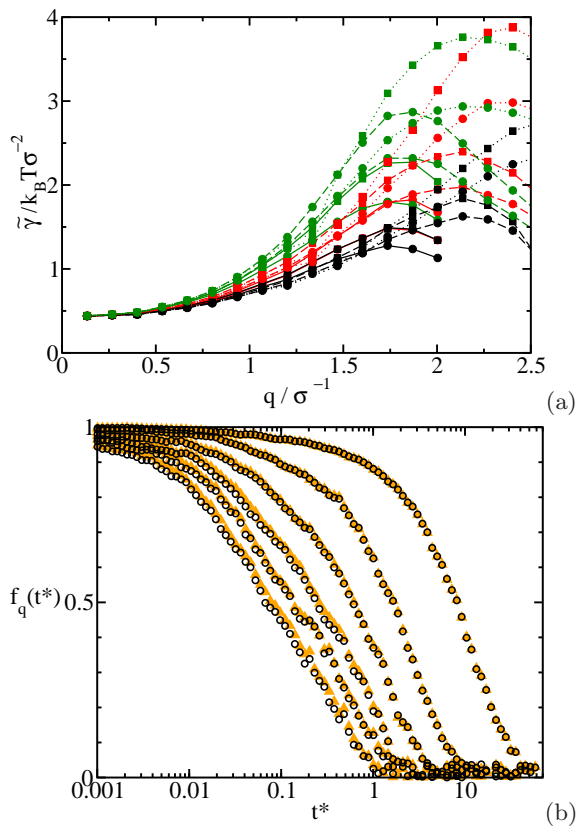

 \includegraphics[width=0.33\paperwidth,height=0.33\paperwidth,keepaspectratio,clip]{loc-interfase-articulo.eps}(a)\\
 \includegraphics[width=0.33\paperwidth,height=0.33\paperwidth,keepaspectratio,clip]{set-params.rescaled.eps}(b)
 \caption{(a) Stiffness as a function of $q$ using different parameters to
locate the interface for the (100)[001] HS system. $\Delta x/\sigma=$ 2.0
(black), 2.5 (red) and 3.0 (green); $n_o=$ 3 (circles) and 4 (squares);  $N=$
30 (continuous line), 40 (dashed line) and 50 (dotted line).  (b)
Autocorrelation function (Eq. \ref{acf}) for the (110)[001] HS system using two different sets of
parameters to locate the interface: $\Delta x=$ 3$\sigma$, $n_0=$ 4 and $N=$ 50 (black circles)
and $\Delta x=$ 2$\sigma$, $n_0=$ 3 and $N=$ 30 (orange triangles). From right
 to left wavevectors $q=2\pi k/L_x$, with $k \le 6$.
} \label{locate_interface}
\end{figure}

The adjustable parameters to obtain a discretized profile of the interface in
the way above described are $N$, $\Delta x$, and $n_o$.  To assess the extent to which the
choice of these parameters affects our results we calculate the interfacial
stiffness via Eq. \ref{eqstiffness} for the HS system using different sets of parameters.  As shown in
Fig. \ref{locate_interface} (a) the stiffness is independent on the 
parameter set for small $q$s, as we approach the 
thermodynamic limit. By contrast, for large $q$s (short wave lengths)
$\widetilde \gamma$ becomes dependent on the analysis parameters.
This dependence is a consequence of the fact that the length scale of the
waves becomes comparable to that of the discretization grid for large $q$s. 
There are sophisticated ways of dealing with this issue \cite{chacon03}, but 
for our purpose it is enough  to stick to the $q$ range where $\widetilde \gamma$
is independent of the analysis parameters (i. e., the six smallest wave vectors). 
%Moreover, large-$q$ CW  
%lack of any physical meaning. 
%For instance, $q$s larger than 3.14 $\sigma^{-1}$ correspond to wave lengths smaller
%than {\em two} particle diameters.  

We also analyse the influence of the analysis parameters on 
the evaluation of the autocorrelation functions. 
In Fig. \ref{locate_interface} (b) we compare $f_q(t)$ 
for the set of analysis parameters used in the main text 
($\Delta x=$ 3$\sigma$, $n_0=$ 4 and $N=$ 50, black circles) and a 
completely different one ($\Delta x=$ 2$\sigma$, $n_0=$ 3 and $N=$ 30, orange
triangles).
Both sets of parameters give virtually identical $f_q(t)$s for the
range of $q$s for which the dynamics has been investigated in this work 
($q = 2\pi k/L_x$ for $k \le 6$).  

\begin{figure}
 \includegraphics[width=0.33\paperwidth,height=0.33\paperwidth,keepaspectratio,clip]{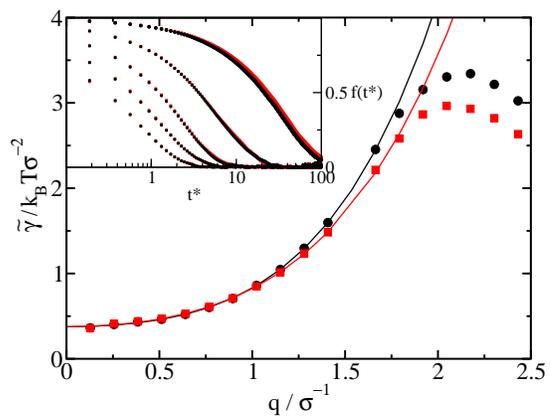}\\
\caption{\textcolor{black}{Main figure: stiffness of the 
(100)[001] LJ interface as a function of $q$ for two different order parameters 
to distinguish between solid and liquid-like particles. Black circles 
correspond to the order parameter described in Appendix \ref{ordparam} and red squares correspond to that described 
in Ref.\cite{peterRoySoc2009}. Inset: Autocorrelation functions obtained with both order parameters
for different wave vectors.}}
\label{efectoparametro}
\end{figure}

\textcolor{black}{We have also analysed the effect  
that the order parameter to distinguish between crystal and fluid-like particles has in our results. 
In the main part of the paper we use for the LJ system the order parameter proposed in Ref.  
 \cite{lechnerDellago08} with the specific parameter set given in Appendix \ref{ordparam}, whereas for the HS system we use 
the order parameter proposed in Ref. \cite{JCP_1996_104_09932} with the parameter
set given in Ref. \cite{peterRoySoc2009}. Both order parameters are inspired by 
Ref. \cite{PhysRevB.28.784} and are devised to distinguish an fcc lattice from the fluid, 
but the order parameter of Ref. \cite{lechnerDellago08} gives more importance to second nearest neighbors
than that of Ref. \cite{JCP_1996_104_09932}.  
Here, we 
recalculate the stiffness and the autocorrelation function of the LJ system with the order
parameter and set of parameters used for the HS sytem. 
The comparison of the results obtained with both order parameters is shown in Fig. \ref{efectoparametro}. Reassuringly, the results are not sensitive
to the specific choice of the order parameter, provided, obviously, that the chosen order parameter is able to distinguish between 
the solid lattice (fcc in this case) and the fluid.  
}

\subsubsection{System size effects}

In order to avoid simulating systems with a prohibitively large number of molecules
and yet be able to probe small-$q$ CW we use simulation boxes with one side 
significantly shorter than the others (see Fig. \ref{snapshot}). This choice results 
in an elongated interfacial area. As shown in Ref.\cite{0295-5075-93-2-26006} it
is advisable to check if the geometry of the simulation box introduces spurious
effects in the surface wave dynamics. This is indeed a potential source of
concern, since the very elongated systems employed are quasi 1--dimensional, and
the behaviour of CW strongly depends on dimensionality.

In this section we provide
some evidence that our choice for the shape of the simulation box does not 
affect our results. 

\begin{figure}
\includegraphics[width=0.33\paperwidth,height=0.33\paperwidth,keepaspectratio,clip]{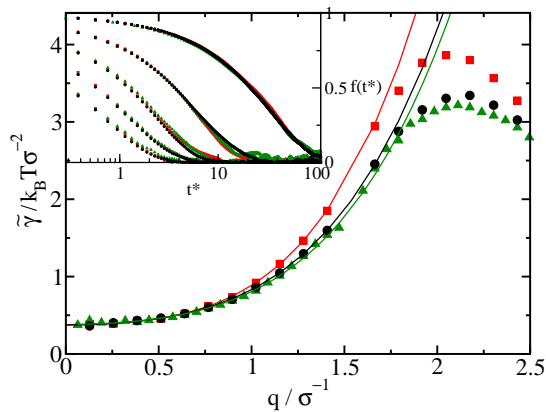}
\caption{Main figure: effect of the interface area in the stiffness. Black circles represent
the stiffness for the (100)[001] LJ system with L$_x$=49.101$\sigma$ 
and L$_y$=6.336$\sigma$. Green triangles and red squares represent the stiffness for the same system 
but with L$_x$'=2L$_x$ and L$_y$'=3L$_y$ respectively. 
Inset: Autocorrelation function for the three systems and different wavevectors.}
\label{tamano}
\end{figure}

To check if the typical system size we use in this work yields system-size-dependent
results we take the (100)[001] LJ system and compare the results
for the size reported in Table \ref{systems_size} (black circles in Fig. \ref{tamano}) with those obtained by making
the short edge of the interfacial area ($y$ axis) three times longer (red squares in Fig. \ref{tamano}). 
We compare both the stiffness (main Fig. \ref{tamano}) and the 
dynamic autocorrelation function (inset Fig. \ref{tamano}) for different wavevectors.  
Clearly, the typical size for the short axis of the simulation box used in this work 
causes no significant finite size effects for the $q$-range we have considered 
for the analysis of the dynamics ($q = 2\pi k/L_x$ for $k \le 6$). 
In Fig. \ref{tamano} we also compare with the results obtained by 
doubling the long edge ($x$-axis) of the interfacial area (green curve). 
Again, no significant finite size effects are seen.  

It is important to note, however, that in order to avoid system size artifacts,
one must study the dynamics of the laterally averaged interface positions
$h(x_n)$. On the contrary, studying the dynamics of the stripes $h(x_n,y_p)$
provides results that are strongly system size dependent.

\subsubsection{Consistency with previous results}

\begin{figure}
\includegraphics[width=0.33\paperwidth,height=0.33\paperwidth,keepaspectratio,clip]{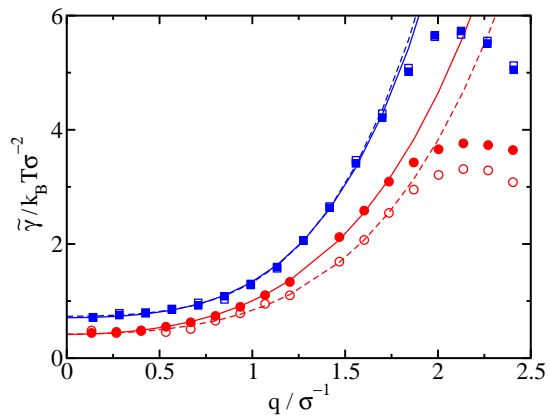}
\caption{Stiffness as a function of $q$ for the HS system and two different orientations: 
(100)[001] (circles) and (110)[001] (squares). 
MC and MD results are shown with open and filled symbols respectively. 
The extrapolation of $\widetilde \gamma (q)$ to $q=0$ in order obtain $\widetilde \gamma$ 
was made by fitting $\widetilde \gamma (q)$ to 
$\widetilde \gamma +aq^2+bq^4$ for small $q$s, where $a$ and $b$ are fitting parameters. 
The fits to MC and MD results are shown as dashed and solid lines respectively. 
} \label{stiffness} \end{figure}

In the discussion above we have already shown that our results are consistent with previous
studies. For instance, in Fig. \ref{compamini}(a) we show that we obtain the same kinetic coefficient
as in Ref. \cite{amini06} for HS. Moreover, in Fig. \ref{growthrate} we show
that our estimate for the rate of crystal growth of ice is in good agreement with 
Ref. \cite{kusalik2011}, where this quantity is calculated through a completely different approach. 
To further validate our methodology we show our results for the interfacial stiffness $\widetilde \gamma$ 
by means of Eq. \ref{eqstiffness}
for two different orientations of the HS system (See Fig.
\ref{stiffness}) and compare it with previously reported values \cite{Davidchack06,Oettel12}. 
The comparison is
shown in Table \ref{comparison_stiffness}.
Our results 
are in good agreement with the literature. Moreover we obtain, as expected for an 
equilibrium property as $\widetilde \gamma$, a good agreement between MC and MD. 
Therefore, 
the way in which we simulate and analyze the interface gives results 
for $\widetilde \gamma$ that are consistent with previously published values. 

\begin{table}
\footnotesize
\begin{tabular}{c c c c c }
 \hline
 \hline
 Orientation & MC & MD & Ref.\cite{Davidchack06} & Ref.\cite{Oettel12} \\
\hline
 (100)[001] & 0.42(2) & 0.415(5) & 0.44(3) & 0.419(5) \\
 (110)[001] & 0.73(2) & 0.707(4) & 0.70(3) & 0.769(5) \\
 \hline
 \hline
\end{tabular}
\caption{Comparison of $\widetilde \gamma$ (in $k_BT/\sigma^2$) obtained in this work 
by means of two different simulation methodologies (MC and MD) with that obtained in previous works for
the HS system and two different orientations.}
\label{comparison_stiffness}
 
\end{table}

\section{Summary and conclusions} 
In this paper we present a computer simulation study of the crystal-melt
interface for three different systems: hard spheres, Lennard Jones and the
TIP4P/2005 model of water. We focus on the dynamics of surface waves. First,
we generate an initial configuration in which a crystal slab is surrounded by
its melt. The box geometry allows for the study of long wave-length capillary
waves without having a prohibitively large number of molecules in the system
(see Fig. \ref{snapshot} for an example).  Then, we do molecular dynamics simulations
in the $NVT$ ensemble at the melting temperature. The overall density of the
system is comprised in between the coexistence densities of the fluid and the
crystal phases, which guarantees that the system stays at coexistence throughout
the $NVT$ simulation. The area of the box side parallel to the interface is
chosen in such way that the solid phase is free of any stress. 

Once we run the molecular dynamics simulations, we analyse the dynamic
autocorrelation function of the surface waves modes (Eq. \ref{acf}). To do that
we first obtain a function that describes the profile of the interface, which
we do by identifying the outermost crystalline particles of the solid slab. 

We carefully checked that our conclusions are not affected by the choice of the 
parameters needed to locate the interface (Fig. \ref{locate_interface}) or by the geometry 
of the box or the system size (Fig. \ref{tamano}).

We examine in detail the shape of the dynamic autocorrelation function as a
function of the wave vector $q$, and conclude that a double exponential function
describes the relaxation dynamics of crystal-melt surface
waves much more accurately than a single exponential (Fig. \ref{ajustes} (a)). 
This implies that there are two distinct time scales, 
fast and slow, involved in the relaxation of crystal-melt surface
waves.  
The
slow time scale is due to the recrystallization-melting occurring at the interface, and
is governed by capillary forces.
The fast relaxation is due to a combination of processes that readily alter the shape
of the interface. We speculate these may be related to Rayleigh waves, 
subdiffusion of the fluid and the attachment/detachment of particles to/from 
the crystal phase. 
As the length scale of the capillary
wave modes increases (or $q$ decreases) the relaxation becomes increasingly
dominated by the slow process and can be  just described by a single
exponential.  Within the uncertainty of our data, we see that the
characteristic time for the slow relaxation process is related to $q$ by the
power law: $\tau \propto q^{-2}$ for all systems.
\textcolor{black}{Note that the conclusions obtained for the surface dynamics of the
hard sphere system, which is not affected by posible artifacts from the
thermsotat,  fully agrees with results for the LJ and water models,
where we resorted to thermostated dynamics.} 
This power law was predicted theoretically in Ref.
\cite{karma93} and checked in simulations of metallic systems \cite{hoyt02,hoyt03,Hoyt20101382,monk2010}
and hard spheres \cite{amini06}, although the existence of a single relaxation process was assumed in these works.
Our results for hard spheres 
are clearly at odds with claims of a linear
overdamped regime observed in the crystal--melt interface of colloids
\cite{Weeks09}.
 
In addition to molecular dynamics simulations, we also perform Monte Carlo
simulations for the hard sphere system. Monte Carlo and molecular dynamics
simulations yield virtually identical autocorrelation functions if compared in
diffusive time units (Fig. \ref{correl-MC_vs_MD}).  \textcolor{black}{This implies 
that the microscopic dynamics
do not play any significant role in the relaxation of CW}. Moreover, we compare the
relaxation dynamics of different systems in diffusive time units. We see that
the crystal-melt interface of water relaxes about ten times slower than that of hard
spheres or Lennard Jones. We ascribe this difference to the
presence of orientational degrees of freedom in the water molecules.     

Following the methodology proposed in Refs. \cite{karma93,hoyt02} we obtain
estimates of the kinetic coefficient (the proportionality constant between the
rate of crystal growth and the supercooling) for all the three systems
investigated. We find a good agreement with the results of Ref. \cite{amini06}
for hard spheres (Fig. \ref{compamini} (a)). In our reduced units we can compare the kinetic
coefficient for Lennard Jones with that of water. We show that a
Lennard Jones crystal grows roughly ten times faster than a water crystal for
the same degree of supercooling.  From the kinetic coefficient we estimate the
rate of crystal growth for ice at moderate supercooling.  We compare it with
recent calculations of such quantity obtained by a completely different
approach \cite{kusalik2011,berne2014} and get a quite good agreement (Fig.
\ref{growthrate}).  However, the linear dependence of the rate of crystal
growth with the supercooling is restricted to fairly small supercooling.  
We also compare our results for the rate of ice growth with experimental
measurements \cite{pruppacher1967} and show that the employed water model predicts
significantly faster rates than those seen in the
experiments (Fig. \ref{growthrate}).

{\bf Acknowledgements}\\
We would like to thank useful discussions with R. P. Sear, R. E. Rozas, J.
Horbach, A. Mijailovic, E. Romero--Enrique, \textcolor{black}{R. Benjamin}, and F. Monroy. 
E. Sanz and J. Benet acknowledge financial support from the EU grant 322326-COSAAC-FP7-PEOPLE-2012-CIG
and from a Spanish grant Ramon y Cajal. L.G. MacDowell and J. Benet also
acknowledge financial support from project FIS2010-22047-C05-05 (Ministerio de
Economia y Competitividad).

%\bibliographystyle{jcp}
%\bibliography{biblio-HS}

\clearpage
\appendix

\section{Order parameters}
\label{ordparam}
To distinguish between solid and liquid-like particles for the LJ and water systems we calculate for each particle 
a local bond order parameter, $\bar q_l(i)$, proposed by Lechner and Dellago \cite{lechnerDellago08}. 
If $\bar q_l(i)$ exceeds a certain threshold particle $i$ is considered to be solid-like. 
The expression for $\bar q_l(i)$ reads:
\begin{equation}
 \bar q_l(i)=\sqrt{\frac{4\pi}{2l+1}\overset{l}{\underset{m=-l}{\sum}}|\bar q_{lm}(i)|^2}, 
\end{equation}
where
\begin{equation}
 \bar q_{lm}=\frac{1}{\widetilde N(i)} \overset{\widetilde N(i)}{\underset{j=1}{\sum}}q_{lm}\left(j\right), 
\end{equation}
and 
\begin{equation}
 q_{lm}=\frac{1}{N_n(i)} \overset{N_n(i)}{\underset{j=1}{\sum}}Y_{lm}\left(r_{ij}\right). 
\end{equation}
Here $\widetilde N$ includes particle $i$ plus all its $N_n$ neighbours,   
and $Y_{lm}$ are $m^{th}$ order spherical harmonics. The neighbors are identified 
over a cut-off distance 
of 3.5 \AA\ for water and 1.4 $\sigma$ for LJ. 

\begin{figure}
 \includegraphics[width=0.33\paperwidth,height=0.33\paperwidth,keepaspectratio]{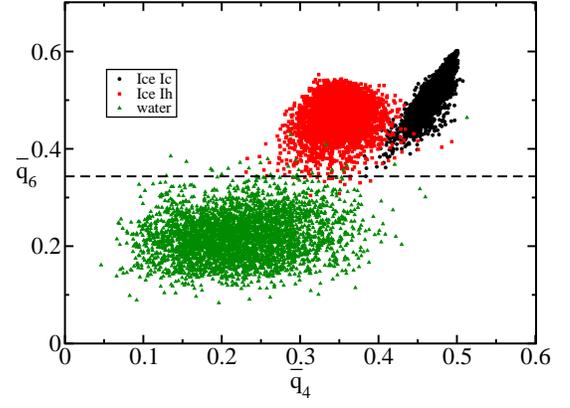}
 \caption{Bidimensional representation of the local bond order parameters for different liquid water (green triangles), ice Ih (red squares)
 and ice Ic (black circles). The corresponding thermodynamic state was T=250K and p=1bar. }
 \label{q6q4}
\end{figure}

\begin{figure}
 \includegraphics[width=0.33\paperwidth,height=0.33\paperwidth,keepaspectratio]{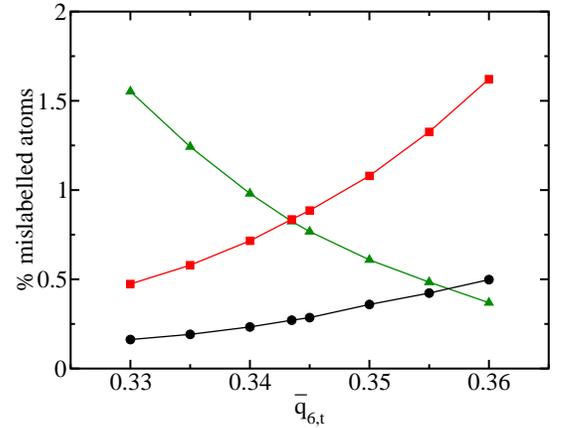}
 \caption{Fraction of atoms mislabelled as a function of $\bar q_{6,t}$. Liquid water (green triangles), 
 ice Ih (red squares) and ice Ic (black circles). }
 \label{fraction-mislabelled}
\end{figure}

In order to determine the best choice for the order parameters we calculated
the values of two order parameters, namely $\bar q_4$ and $\bar q_6$, for both the bulk solid and the bulk fluid
phases at coexistence. We discuss here the case of water. 
As it can be seen in Fig. \ref{q6q4} $\bar q_6$ allows for a better separation between the solids (hexagonal and cubic ice) and the fluid phase in water. 
Next, to choose the $\bar q_6$ threshold ($\bar{q}_{6,t}$)
that best separates the liquid from the solids, we count the percentage of mislabelled molecules in each phase for several choices
of $\bar{q}_{6,t}$. Whenever a liquid particle has a $\bar{q_6}$ value larger than $\bar{q}_{6,t}$, it will be mislabelled as
solid-like. Likewise, if a solid-like particle happens to have a $\bar{q_6}$ smaller than $\bar{q}_{6,t}$, it will be mislabelled as
liquid-like. In Fig. \ref{fraction-mislabelled} we plot the percentage of mislabelled molecules as a function of $\bar{q}_{6,t}$. At $\bar{q}_{6,t}=0.3435$
the liquid and ice-Ih curves cross at a mislabeling percentage of 0.82. We choose that value as
the threshold to discriminate between liquid-like and solid-like molecules.
The threshold is indicated with a horizontal dashed line in Fig. \ref{q6q4}. 
At $\bar{q}_{6,t}=0.3435$ the percentage of mislabelled ice-Ic molecules is as low as 0.26.
This means that ice-Ic molecules would be detected as solid-like, should they appear when the interface recrytallizes.
Once molecules are labelled either as solid or as liquid-like, the largest solid cluster
is found using a clustering algorithm with a cut-off of 3.5~\AA\ to find neighbors belonging to the same cluster.

For the LJ system we used a $\bar{q}_{6,t}$ value of 0.294 and a cut-off to build the 
biggest cluster of 1.4$\sigma$. 

For the HS system we employed an order parameter based on $q_6$ as described in Ref. \cite{peterRoySoc2009} to distinguish
between solid-like and liquid-like particles. 

\section{Double expoential fits}
\label{defit}

\begin{figure}
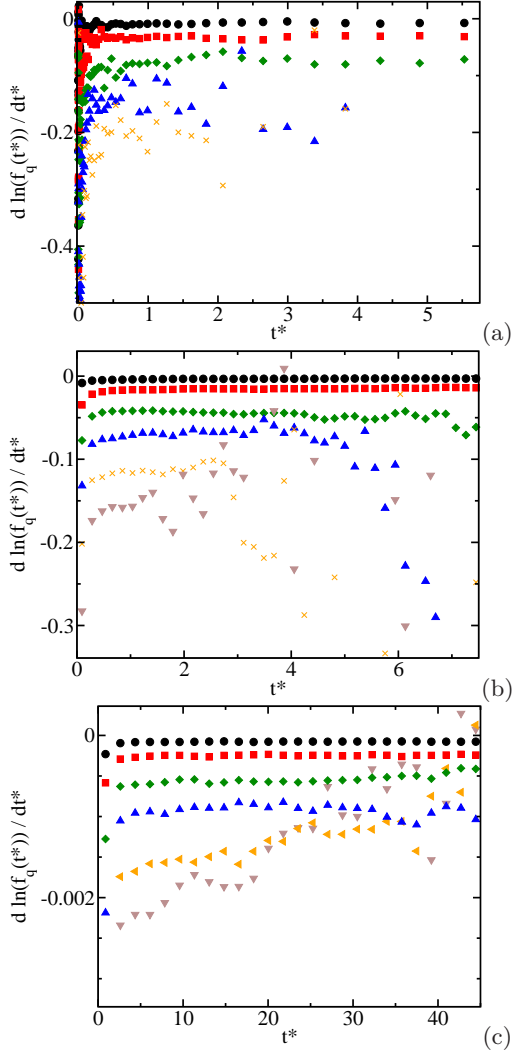


\includegraphics[width=0.35\textwidth,keepaspectratio,clip]{derivadas.hs.100.md.rescaled.eps}(a)
\includegraphics[width=0.35\textwidth,keepaspectratio,clip]{derivadas.lj.100.rescaled.eps}(b)
\includegraphics[width=0.35\textwidth,keepaspectratio,clip]{derivadas.pI.basal.rescaled.eps}(c)
\caption{Representation of $d\ln(f_q(t)/dt$ vs $t$ for three different systems:
(100)[001] HS MD (a), (100)[001] LJ (b) and (pI)[basal] TIP4P/2005 water. From 
top to bottom in a given plot wavevectors with values of $q=2\pi k/L_x$ with $k \le 6$ are shown.}
\label{derivadas}

\end{figure}

When fitting the time autocorrelation functions to a double exponential
function (Eq. \ref{de}) it is convenient to have a good initial guess for the
fitting parameters to avoid  convergence to non-physical results.  In order
to obtain a good guess for the relaxation time of the slow process, $\tau_{ds}$, 
we plot
$d\ln(t)/dt$ vs $t$ (see Fig. \ref{derivadas}). 
Note that for
long enough times 
one can approximate Eq.  \ref{de} by: 
\begin{equation}
f_q(t)\approx Ae^{-t/\tau_{ds}}, 
\label{aprox_1exp}
\end{equation}
taking into account that $\tau_{ds}>>\tau_{df}$.
If we now take the logarithm of Eq. \ref{aprox_1exp} and differenciate with respect 
to $t$ we obtain: 
\begin{equation}
\frac{d \ln(f_q(t))}{dt} \approx \frac{-1}{\tau_{ds}}.
\end{equation}
Therefore, from the 
intercept of the horizontal region of the  plots shown in Fig. \ref{derivadas}  
we can get an estimate of $\tau_{ds}$. Recall that the higher the $q$ the 
more influenced is the relaxation of the interface by the fast process (see Fig. \ref{coefa0}).
Hence, as expected,  the horizontal region in Fig. \ref{derivadas}
becomes less evident as $q$ increases. Nevertheless, it is enough for our purpose of getting an initial
estimate for the fitting parameter $\tau_{ds}$. 

Given that for small $q$s 
the preexponential factor $A$ is close to 1, we use $A = 1$ 
as an initial guess to fit the autocorrelation function corresponding to
the smallest $q$. Regarding $\tau_{df}$, we set an initial value two
orders of magnitude smaller than $\tau_{ds}$.
We use the resulting parameters $A$ and $\tau_{df}$ of the fit to the smallest $q$ as
an input for the following $q$. For $\tau_{ds}$ we use the value estimated
from Fig. \ref{derivadas} as explained above. We repeat this process until
we obtain a fit for each and every $q$.

~\\

\section{$NVE$ vs $NVT$}
\label{nvt-nve}
\begin{figure}

\includegraphics[width=0.35\textwidth,keepaspectratio,clip]{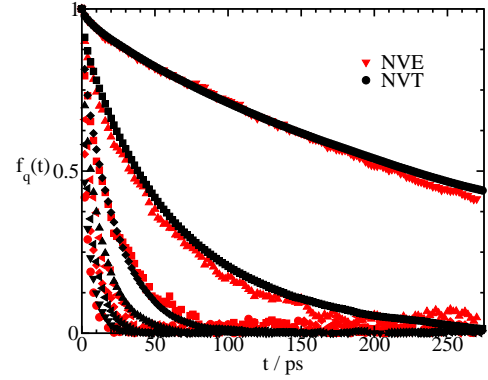}
\caption{Autocorrelation functions for the (100)[001] LJ system as calculated in the $NVT$ and 
$NVE$ ensembles.}
\label{fig-nvt-nve}

\end{figure}

In Fig.\ref{fig-nvt-nve} we compare the autocorrelation functions calculated in many short $NVE$ simulations 
starting from independent configurations with
those obtained in a single, long $NVT$ simulation. $NVE$ simulations are short enough to guarantee energy conservation
and, at the same time, long enough to allow for the relaxation of the studied capillary waves modes. Clearly, $NVE$
and $NVT$ give the same results ($NVE$ curves are more noisy because the statistics is not as good).

\end{document}